\documentclass{ws-procs975x65}

\usepackage{graphicx}

\newcommand\beq{\begin{equation}}
\newcommand\eeq{\end{equation}}
\newcommand\beqa{\begin{eqnarray}}
\newcommand\eeqa{\end{eqnarray}}

\newcommand{\ds}[1]{#1 \hspace{-0.5em}/}  
\newcommand\bzeta{\mbox{\boldmath$\zeta$}}

\newcommand\bsigma{\mbox{\boldmath$\sigma$}}
\newcommand\bSigma{\mbox{\boldmath$\Sigma$}}

\newcommand\bk{{\bf k}}
\newcommand\bq{{\bf q}}

\newcommand\bphi{\mbox{\boldmath$\phi$}}
\newcommand\bp{{\bf p}}

\begin{document}

\title{Ferromagnetism in QCD phase diagram}

\author{T. Tatsumi}

\address{Department of Physics, Kyoto University,
Kyoto 606-8502, Japan\\
E-mail: tatsumi@ruby.scphys.kyoto-u.ac.jp}

\begin{abstract}
A possibility and properties of spontaneous magnetization in quark
matter are 
investigated. Magnetic susceptibility is evaluated within Fermi liquid theory, taking into account
of screening effect of gluons. Spin wave in the polarized quark matter, as the Nambu-Goldstone mode,
is formulated by way of the coherent-state path integral.
\end{abstract}


\bodymatter

\section{Introduction}\label{aba:sec1}

The phase diagram of QCD has been elaborately studied in
density-temperature plane. Here we study the magnetic properties of
QCD. Since the discovery of magnetars with super-strong magnetic field of 
$O(10^{14-15})$G the origin of strong magnetic
field observed in compact stars has roused our interest again \cite{har}. For the
present there are three ideas about its origin: the first one reduces it to the
fossil field, assuming the conservation of the magnetic flux during the
evolution of stars. The second one applies the dynamo mechanism in the crust
region. The third one seeks it at the core region, where hadron or
quark matter develops. If spins of nucleons or quarks are aligned in
some situations, they can provide the large magnetic field. Since
elaborate studies about the spontaneous magnetization in nuclear matter
have repeatedly shown the negative results, in the following,  
we explore the possibility of the spontaneous magnetization in quark
matter \cite{tat00,nak03}. 

In the first paper \cite{tat00} we have suggested a possibility of
ferromagnetic phase in QCD, within the one-gluon-exchange (OGE)
interaction. If such a phase is realized inside compact stars, we can
see the magnetic field of $O(10^{15-17})$G is easily obtained. 
Directly evaluating the total energy of the polarized 
matter with a polarization parameter, we have shown that ferromagnetic
phase is possibly realized at low densities, in analogy with the
itinerant electrons by Bloch \cite{blo}; the Fock exchange interaction is
responsible to ferromagnetism in QCD. We have also seen that the phase
transition is weakly first order and one may also apply 
the technique for the second order phase transition to analyze it, while
it is one of the specific features of the magnetic transition in gauge
theories.

In the first half we discuss the magnetic susceptibility of
quark matter within the Fermi-liquid theory \cite{bay04}. 
It is well known that we
must properly take into account the screening effects in the gluon
propagator to improve the IR behavior of the gauge interaction. For the
longitudinal gluons we can see the static screening described in terms of
the Debye screening mass. There is no static screening for the
transverse gluons, while the dynamical screening appears instead due to
the Landau damping \cite{sch99}. We will figure out these screening effects in
evaluating the magnetic susceptibility. It would be interesting to
observe that there appears the non-Fermi-liquid effect to give an
anomalous term in the finite temperature case as in the specific heat \cite{hol}.

In the second half we discuss how the spin wave, which is caused by the
spontaneous magnetization, can be described within the coherent-state
path integral \cite{rad}. Fist we map the quark matter to a spin system by assuming
the spatial wave function for each quark and leaving the degree of
freedom of the direction of the spin vector. This method is inspired by
the spiral approach taken by Herring in old days to discuss the spin
wave in the electron gas \cite{her66}. Introducing the collective variables ${\bf
U}(\bar\theta,\bar\phi) \in S^2$, and integrating over the individual variables,
we have an effective action to see the classical Landau-Lifshitz
equation for the spin wave is naturally derived by the effective action
for the collective variables. We also note that there are some
geometrical aspects in the effective action, which may further quantize
the classical spin wave to give magnons. Thus we have magnons in the
ferromagnetic phase of quarks, which directly gives the $T^{2/3}$
dependence for the reduction of the magnetization.

\section{Magnetic susceptibility within the Fermi liquid theory}

The magnetic susceptibility $\chi_M$ is defined as
\beq
\chi_M=\left.\frac{\partial\langle M\rangle }{\partial B}\right|_{N,T},
\eeq
with the magnetization $\langle M\rangle$.
So we study the response of quark matter when a weak magnetic field is
applied for a given quark number $N$ and temperature $T$.
Using the Gordon identity, the interaction Lagrangian may be written 
for the constant magnetic field,
${\bf A}={\bf B}\times{\bf r}/2$;
\beq
\int d^4x{\cal L}_{ext}=\mu_q\int d^4x\bar q\left[-i{\bf
r}\times\nabla+\bSigma\right]\cdot{\bf B}q,
\eeq
for quarks with electric charge $e_q$, where $\bSigma=\left(\begin{array}{@{\,}cccc@{\,}}
\bsigma & 0\\0 & \bsigma\end{array}\right)$, and $\mu_q$ denotes the Dirac magnetic moment $\mu_q=e_q/2m$.
Then magnetization $\bf M$ may be written as 
\beq
M_z=\langle
\bar q\bSigma q\rangle_z=\frac{\mu_q}{2}N_C\int\frac{d^3k}{(2\pi)^3}g_D(\bk)(n_{\bk, +}-n_{\bk,-}),
\eeq
with the Fermi-Dirac distribution function $n_{\bk,\zeta}$ in the presence of $\bf B$.
The gyromagnetic ratio $g_D$ is defined as 
\beq
g_D(\bk)\zeta=2{\rm
tr}\left[\Sigma_z\rho(k,\zeta)\right]=\left[1-\frac{k_z^2}{E_k(E_k+m)}\right]\zeta,
\eeq
in terms of the polarization density
matrix $\rho(k,\zeta)$,
\beq
\rho(k,\zeta)=\frac{1}{2m}(\ds{k}+m)P(a),
\eeq
with the projection operator, $P(a)=1/2\cdot(1+\gamma_5\ds{a})$ \cite{}.

The quasi-particle interaction consists of two terms,
\beq
f_{\bk\zeta,\bq\zeta'}=f_{\bk,\bq}^s+\zeta\zeta'f_{\bk,\bq}^a,
\eeq
where $f_{\bk,\bq}^{s(a)}$ is the spin-independent (dependent)
interaction.  
Then we get the expression for the magnetic susceptibility at $T=0$ written in terms of 
the Landau parameters:
\beq
\chi_M=\left(g_D^F\mu_q/2\right)^2\left[\frac{\pi^2}{N_Ck_F\mu}-\frac{1}{3}f_1^s+\bar
f^a\right]^{-1},
\eeq
where $f_1^s$ is a spin-averaged Landau parameter defined by 
\beq
f_1^s=\left.\frac{3}{4}\sum_{\zeta,\zeta'}\int\frac{d\Omega_{\hat{\bk\bq}}}{4\pi}\cos\theta_{\hat{\bk\bq}} 
f^s_{\bk,\bq}\right|_{|\bk|=|\bq|=k_F},
\eeq
with the relative angle $\theta_{\hat{\bk\bq}}$ of $\bk$ and $\bq$, and $\bar f^a$ the spin-dependent one,
\beq
\bar f^a\equiv \left.\int\frac{d\Omega_\bk}{4\pi}\int\frac{d\Omega_\bq}{4\pi}f^a_{\bk,\bq}\right|_{|\bk|=|\bq|=k_F}.
\eeq

\section{Static and dynamic screening effects}

When we consider the color-symmetric forward scattering amplitude of the
two quarks around the Fermi surface by the one gluon
exchange interaction (OGE), the direct term should be vanished due to
the color neutrality of quark matter and the Fock exchange term gives a
leading contribution.  
The color-symmetric and flavor-symmetric  OGE interaction of quasi-particles on the Fermi surface may be written,
\beq
\left.f_{\bk\zeta,\bq\zeta'}\right|_{|\bk|=|\bq|=k_F}
=\left.\frac{1}{N_C^2}\frac{1}{N_F^2}\sum_{a,b}\sum_{i,j}f_{\bk\zeta
a i,\bq\zeta' b j}\right|_{|\bk|=|\bq|=k_F}
=\left.\frac{m}{E_k}\frac{m}{E_q}M_{\bk\zeta,\bq\zeta'}\right|_{|\bk|=|\bq|=k_F},
\eeq
with the invariant matrix element, $M_{\bk\zeta,\bq\zeta'}$ \cite{bay76}.
It has been well known that massless gluons often causes infrared (IR)    
divergences in the Landau parameters \cite{bay04}.

Since the one gluon exchange interaction is a long-range force and we
consider the small energy-momentum transfer between quasi-particles, we
must improve the gluon propagator by taking into account the screening effect,
\beq
D_{\mu\nu}(k-q)=P^T_{\mu\nu}D_T(p)+P^L_{\mu\nu}D_L(p)-\xi\frac{p_\mu
p_\nu}{p^4} 
\eeq
with $p=k-q$, where $D_{T(L)}(p)=(p^2-\Pi_{T(L)})^{-1}$, and 
the last term represents the gauge dependence with a parameter
$\xi$. $P^{T(L)}_{\mu\nu}$ is the standard projection operator onto the
transverse (longitudinal) mode.

The self-energies for the transverse and longitudinal gluons are given
as 
\beqa
\Pi_L(p_0,{\bf p})&=&m_D^2+i\frac{\pi
m_D^2}{2v_F}\frac{p_0}{|\bp|}{\rm coth}\frac{p_0}{2T}\nonumber\\
\Pi_T(p_0,{\bf p})&=&-i\frac{\pi p_F m_D^2}{4E_F}\frac{p_0}{|\bp|}{\rm coth}\frac{p_0}{2T}, 
\eeqa
in the limit $p_0/|\bp|\rightarrow 0$, with the Debye screening mass,
$m_D^2\equiv (N_F/2\pi^2)g^2 E_F k_F$ \cite{}. 
Thus the longitudinal gluon is
screened to have the Debye mass $m_D$, while the transverse gluon is not
in the limit $p_0/|{\bf p}|\rightarrow 0$ and $T=0$ \cite{sch99}. At finite
temperature, however, the transverse gluons are also dynamically
screened by the Landau damping.
For quarks on the Fermi surface, the Lorentz invariant matrix element can be written as 
\beq
M_{\bk\zeta,\bq\zeta'}=-\frac{N_C^2-1}{2N_C^2}g^2\left[-M^{00}D_L+M^{ij}P_T^{ij}D_T
+\xi\frac{1}{|\bk-\bq|^4}(\bk-\bq)_i(\bk-\bq)_jM^{ij}\right],
\eeq
with the coefficients $M^{\mu\nu}$, 
\beqa
M^{\mu\nu}&=&{\rm
tr}[\gamma^\mu\rho(k,\zeta)\gamma^\nu\rho(q,\zeta')].
\eeqa 
 First of
all, the matrix element is obviously independent of the gauge parameter
$\xi$.

At temperature $T=0$, there is no screening in the propagator of the
transverse gluon, so that logarithmic divergence still remains in the
Landau-Migdal parameters, $f_1^s$ or $\bar f^a$. However, we can see
that the divergences
cancels each other to give a finite result for the susceptibility. The
static screening effect gives a $g^2\ln g^{-2}$ contribution \cite{tatsat}.

\begin{figure}[h]
\begin{center}
\includegraphics[height=5.cm]{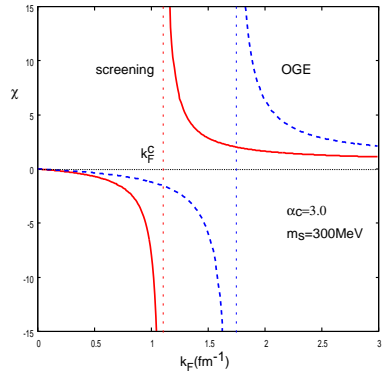}
\caption{
Magnetic susceptibility as a function of the Fermi momentum. Divergence signals
 the onset of spontaneous spin polarization. The screening effect
 slightly shifts the critical density to lower densities.
}
\label{f:mix}
\end{center}
\end{figure}

In Fig. 1 we present an example of the magnetic susceptibility at
$T=0$. The ferromagnetic phase corresponds to the negative value of
$\chi$ and the critical density is given by the divergence of $\chi$.
Compared with the OGE calculation, we can see that the static
screening effect shifts the critical density to the lower value.

At finite temperature, the dynamical screening gives $T^2\ln T$
contribution through the density of states near the Fermi surface,
besides the standard $T^2$ dependence. This behavior is a kind of
non-Fermi liquid effects, as in the specific heat \cite{hol}.

\section{Spin wave in the polarized quark matter}

When spontaneous magnetization occurs, the magnetization ${\bf M}$ is
not vanished, so that rotation symmetry is violated in the ground
state. Hence one can expect a Nambu-Goldstone mode, {\it spin wave}
there. Different from the usual description of the spin wave, we must
care about its realization in quark matter. Recalling a similar
situation in electron gas, we take here an intuitive but correct
framework called the spiral approach. Herring explicitly showed that the Bloch wall
coefficient, which is closely related with the dispersion relation of
the spin wave, is obtained for electron gas by the use of the spiral
approach \cite{her66}. 

Consider the fully polarized case. Then all the quarks have a definite
spin state specified by 
$\bzeta$ (say, $\zeta=+1$). The single particle wave function with
momentum $k$ is simply given as
\beq
u^{(\zeta)}(k)=\frac{\ds{k}+m}{\sqrt{2m(E_k+m)}}u^{(\zeta)}(m,{\bf 0})e^{-ikx}
\eeq
with the spinor in the rest frame, 
\beq
u^{(\zeta=+1)}(m,{\bf 0})=e^{ib}
\left(
\begin{array}{c}
e^{-i\phi_k/2}\cos\theta_k/2\\
e^{i\phi_k/2}\sin\theta_k/2\\
0
\end{array}
\right)\equiv
\left(
\begin{array}{c}
 g_k({\bf r})\\
0
\end{array}
\right)
\eeq
taking the spin quantization axis specified by the polar angles
$(\theta_k, \phi_k)$. Thus the quark wave function is characterized by the momentum $\bf k$
and the polar variables  $\theta_k,\phi_k$. 

In the ground state, all the spins have the same direction, say
$\theta_k=\bar\theta,\phi_k=\bar\phi$, so that 
the ground-state energy is degenerate in any value of them. In the following
we allow them to be spatially dependent and introducing the small
fluctuation fields, $\xi_k(x),\eta_k(x)$, s.t.
\beq
\theta_k(x)=\bar\theta(x)+\xi_k(x),   \phi_k(x)=\bar\phi(x)+\eta_k(x),
\eeq 
where $\bar\theta,\bar\phi$ or ${\bf U}=(U^1(\bar\theta,\bar\phi),
U^2(\bar\theta,\bar\phi),U^3(\bar\theta,\bar\phi))$ are the collective
variables defined by 
\beq
\cos\bar\theta\equiv\frac{1}{N_k}\sum_k\cos\theta_k, \bar\phi\equiv\frac{1}{N_k}\sum_k\phi_k,
\eeq
and we shall
see they describe the Nambu-Goldstone mode (spin wave). 


In the spiral approach, we assume a spin configuration described by  
\beq
g_k({\bf r})=\exp\left[it_z(d\bar\phi/dz)(\sigma_z/2)\right]g_k({\bf r-t})
\eeq
with arbitrary displacement ${\bf t}$, which corresponds to a spin wave
traveling along $z$ axis with wave vector $d\bar\phi/dz$, so that 
$\bar\phi$ should be a linear function of $z$. Actually the mean 
value of the spin operator $\bSigma$ is proportional to $\bar\bzeta$.

For a given Hamiltonian $H$ we can evaluate the energy by putting (15),
(16) into it,
\beq
H(\theta_k,\phi_k)=\int d^3x\langle H\rangle,
\eeq
which may be regarded as a classical spin Hamiltonian for quark
matter. Thus we mapped quark matter to assembly of spins.

\section{Coherent-state path-integral}

One may reformulate the idea of the spiral approach in the framework of
the path integral \cite{rad}. 
Consider a matrix element of the evolution operator 
\beq
\langle\Omega'',t''\left|\right.\exp(-iT\hat H/\hbar)\left.\right|\Omega',t'\rangle
=\int{\cal D}\Omega\exp\left(i\sum_k\int_{t'}^{t''}dt\int d^3x\left[i\langle\Omega_k\left.\right|\dot\Omega_k\rangle-H(\Omega)\right]\right)
,
\eeq
where $|\Omega\rangle=|\Omega_1\rangle\otimes\cdot\cdot\cdot\otimes|\Omega_{N_k}\rangle$ is the spin
coherent state 
\beq
\left|\Omega_k\rangle\right.=\left(\cos(\theta_k/2)\right)^{2S}\exp\left[\tan\left(\theta_k/2\right)e^{i\phi_k}\hat
S_-\right]|0\rangle,
\eeq
with $S_z|0\rangle=S|0\rangle$.

Introducing collective variables $\bar \Omega=(\bar\theta,\bar\phi)$, 
Eq.~(21) may be rewritten as
\beqa
&&\int{\cal D}\Omega{\cal
D}\bar\Omega\delta\left(\bar\Omega-\frac{1}{N_k}\sum\Omega_k\right)
e^{\left\{\frac{i}{\hbar}\sum_k\int_{t'}^{t''}dt\int d^3x\left[\frac{1}{2}(1-\cos\theta_k)\dot\phi_k-H(\theta_k,\phi_k)\right]\right\}}\nonumber\\
&=&\int{\cal D}\bar\Omega e^{iS_{\rm eff}(\bar\Omega)},
\eeqa
where the {\it effective action} is defined as 
\beq
e^{iS_{\rm eff}(\bar\Omega)}=\int{\cal D}\xi{\cal
D}\eta\delta\left(\sum_k\xi_k\right)\delta\left(\sum_k\eta_k\right)
e^{\left\{\frac{i}{\hbar}\sum_k\int_{t'}^{t''}\int d^3x\left[\frac{1}{2}(1-\cos(\bar\theta+\xi_k))(\dot{\bar
\phi}+\dot\eta_k)-H(\bar\theta,\bar\phi,\xi_k,\eta_k)\right]dt\right\}}.
\eeq
We expand the exponent with respect to $\xi_k,\eta_k$ up to the
second-order, discarding the higher-order terms in $\dot{\bar\phi}$ within 
the adiabatic approximation.
Taking the stationary-phase approximation s.t.
\beq
\left.\frac{\delta H(\bar\theta,\bar\phi,\xi_k,\eta_k) }{\delta\xi_k}\right|_{\xi_k=\xi_k^c}=\left.\frac{\delta H(\bar\theta,\bar\phi,\xi_k,\eta_k)}{\delta\eta_k}\right|_{\eta_k=\eta_k^c}=0,
\eeq
we have
\beq
S_{\rm eff}\approx \frac{i}{\hbar}\int_{t'}^{t''}dt\int d^3x
\left[\Sigma(1-\cos\bar\theta)\dot{\bar\phi}-H(\bar\theta,\bar\phi,\xi_k^c,\eta_k^c)\right],
\eeq
with $\Sigma=N_k/2V$.
One may show that
\beq
H(\bar\theta,\bar\phi,\xi_k^c,\eta_k^c)=\frac{A}{2}\left(\nabla_r{\bf U}\right)^2
\eeq
as should be. Then, the Bloch wall coefficient reads
\beq
A=\frac{N_k/V}{8E_F}+O(g^2).
\eeq
The classical equation of motion for $\bf U$ is given by $\delta S_{\rm
eff}=0$,
\beq
\Sigma \dot{\bf U}+2A\Delta{\bf U}\times{\bf U}=0,
\eeq
which is the classical Landau-Lifshitz equation for the spin wave. Then, the dispersion relation for
the spin wave is deduced,
\beqa
\omega(\bq)&=&(2A/\Sigma)q^2\nonumber\\
&=&\left(1/2E_F\right)q^2+O(g^2).
\eeqa

The effective action has some topological features, with which we can do
the second quantization for the spin wave.
The first term in the effective action can be written as the interaction
with the dynamically induced vector potential $\bf A$:
\beq
(1-\cos\bar\theta)\dot{\bar\phi}={\bf A}({\bar\bzeta})\cdot
{\dot{\bar\bzeta}}    
\eeq
with
\beq
{\bf A}=\left(\frac{1-\cos\bar\theta}{\sin\bar\theta}\right)\hat{\bphi}.
\eeq
Employing the Dirac quantization condition, we can see that
$\Sigma=$integer or half-integer as should be.

Alternatively, the first term in terms of $\bf U$ can be rewritten as
the line integral along the path on $S^2$, which is nothing but the Wess-Zumino term.

From the effective action (26), it is inferred that $\cos\bar\theta$ and
$\bar\phi$ are canonical conjugate. Putting
$
\bar\pi\equiv\Sigma(1-\cos\bar\theta),
$ 
 one can verify
\beq
\left\{\Sigma U^+({\bf x},t), \Sigma U^-({\bf
y},t)\right\}_{PB}=2i\Sigma U^3({\bf x},t)\delta({\bf x}-{\bf y})\simeq 2i\Sigma\delta({\bf x}-{\bf y}),
\eeq
where we have used the Kramers-Heller approximation for a large number
of particles in the last
step. Taking the Fourier transform, s.t.,
\beq
U^+({\bf x},t)=\sqrt{\frac{2}{\Sigma}}\sum_{\bf q}a_{\bf q}e^{i{\bf q}\cdot{\bf x}},
\eeq
and $U^-({\bf x},t)=\left(U^+({\bf x}, t)\right)^\dagger$,
and assuming the quantum-mechanical commutation relation ({\it second quantization}),
\beq
[a_{\bf q},a^\dagger_{\bf q'}]=\delta_{{\bf q},{\bf q'}},
\eeq
we have an Hamiltonian as an assembly of magnons,
\beq
\int d^3x H(\bar\theta,\bar\phi,\xi_k^c,\eta_k^c)=\sum_{\bf q} \omega(\bq)
a_{\bf q}^\dagger a_{\bf q}.
\eeq


For low temperature the thermodynamical properties may be described by
the excitation of the spin wave. We can easily see that magnetization is
reduced by the excitation of the spin waves at finite temperature
$(T^{3/2}~{\rm Law})$,
$(M(T)-M(0))/N_kg_D\mu_q=-\zeta(3/2)\left(\Sigma T/8\pi A\right)^{3/2}$,
from which the Curie temperature reads,
$T_c=8\pi A/\Sigma (\rho/2\zeta(3/2))^{2/3}$. Thus we can roughly estimate
the Curie temperature of several tens MeV for $2A/\Sigma\simeq 1/2E_F$

\section{Concluding remarks}

In this paper we have discussed the critical line of the spontaneous
polarization on the density-temperature plane within the framework of
the Fermi-liquid theory. We have evaluated the magnetic susceptibility
by taking into account the screening effects for the gluon propagation,
and figured out the important roles of the static and dynamic screening;
the former gives the $g^2\ln(1/g^2)$ contribution, while the latter
gives $T^2\ln T$ for finite temperature. Both effects surely reflect
the specific feature of the gauge interaction. To get more realistic values
for the critical density and temperature, we must consider some
nonperturbative effects such as instantons as well as the systematic
analysis of the higher-order terms in QCD.

We have presented a framework to deal with the spin wave as a
Nambu-Goldstone mode. In the perspectives
we can derive the magnon-quark coupling vertex, which may give a novel
cooling mechanism and a novel type of the Cooper pairing \cite{kar}. They are not
only theoretically interesting, but also phenomenologically important
for the thermal evolution of compact stars bearing the ferromagnetic
phase inside. It should be interesting if magnon effects could
distinguish the microscopic origin from fossil field or the dynamo
mechanism. 

The author thanks K. Sato for his collaboration.
This work has been partially supported by the Grant-in-Aid for the
21st Century COE ``Center for the Diversity and Universality in Physics''
and the Grant-in-Aid for Scientific Research Fund (C) of the Ministry of
Education, Culture, Sports, Science and Technology of Japan (16540246). 




\begin{thebibliography}{9}
\bibitem{har} A.K. Harding and D. Lai, Rep. Prog. Phys. 69 (2006) 2631.

\bibitem{tat00} T. Tatsumi, Phys. lett. {\bf B489} (2000) 280.\\
                T. Tatsumi, E. Nakano and K. Nawa, {\it Dark
	Matter} (Nova Science Pub., New York, 2006), 39.

\bibitem{nak03} E. Nakano, T. Maruyama and T. Tatsumi, Phys. Rev. {\bf
	D68} (2003) 105001.\\
                T. Tatsumi, E. Nakano and T. Maruyama,
	Prog. Theor. Phys. Suppl. {\bf 153} (2004) 190.\\
                T. Tatsumi, T. Maruyama and E. Nakano, {\it Superdense
	QCD Matter and Compact Stars} (Springer, 2006), 241.

\bibitem{blo} F. Bloch, Z. Phys. {\bf 57} (1929) 545.

\bibitem{bay04} G. Baym and C.J. Pethick, {\it Landau Fermi-Liquid
	Theory} (WILEY-VCH, 2004)\\
                P. Nozi\'eres, {\it Theory of Interacting Fermi Systems}
	 (Westview Press,1997).\\
                A.A. Abrikosov, L.P. Gorkov and I.Ye. Dzyaloshinskii,
	{\it Quantum Field Theoretical Methods in Statistical Physics}
	(Pergamon, Oxford, 1965).


\bibitem{sch99} M. Le Bellac, {\it Thermal Field Theory} (Cambridge
	U. Press, 1996).\\
                T. Sch\"afer and F. Wilczek, Phys. Rev. {\bf D60} (1999)
	114033.

\bibitem{hol} T. Holstein, R.E. Norton and P. Pincus, Phys. Rev. {\bf
	B8} (1973) 2649.\\
S. Chakravarty, R.E. Norton and O.F. Syljuasen, Phys. Rev. Lett. {\bf
	74} (1995) 1423.\\
A. Gerhold, A. Ipp and A. Rebhan, Phys. Rev. {\bf D70} (2004) 105015.
A. Schaf\"ar and K. Schwenzer, Phys. Rev. {\bf D70} (2004) 054007; 114037.


\bibitem{rad} J.M. Radcliffe, J. Phys. {\bf A4} (1971) 313.\\
J.R. Klauder, Phys. Rev. {\bf D19} (1979) 2349.

\bibitem{her66} C. Herring, Phys. Rev. {\bf 85} (1952) 1003.\\
C. Herring, {\it Exchange Interactions among Itinerant
	Electrons: Magnetism IV} (Academic press, New
	York, 1966).



\bibitem{bay76} G. Baym and S.A. Chin, Nucl. Phys. {\bf A262} (1976)
	527.


\bibitem{tatsat} T. Tatsumi and K. Sato, in preparation. 



\bibitem{kar} N. Karchev, J. Phys.:Condens. Matter 15 (2003) L385.\\
              D. Fay and J. Appel, Phys. Rev. {\bf B22} (1980) 3173.0
\end{thebibliography}
\end{document}